\documentstyle[preprint,aps]{revtex}
\begin{document}
\title{Unusual ground-state properties of the Kondo-lattice compound Yb$_2$Ir$_3$Ge$_5$}
\author{Yogesh Singh, S. Ramakrishnan}
\address{Tata Institute of Fundamental Research, Mumbai-400005, India}
\maketitle
\begin{abstract}
\noindent
We report sample preparation, structure, electrical resistivity, magnetic susceptibility and heat capacity studies of a new 
compound Yb$_2$Ir$_3$Ge$_5$. We find that this compound crystallizes in an orthorhombic structure 
with a space group PMMN unlike the compound Ce$_2$Ir$_3$Ge$_5$ which crystallizes in the tetragonal 
IBAM (U$_2$Co$_3$Si$_5$ type) structure. Our resistivity measurements indicate that the compound Yb$_2$Ir$_3$Ge$_5$ behaves like a typical Kondo lattice system with no ordering down to 0.4~K. However, a Curie-Weiss fit of the inverse magnetic susceptibility above 100~K gives an effective moment of only 3.66 $\mu$$_B$ which is considerably less than the theoretical value of 4.54 $\mu$$_B$ for magnetic Yb$^3+$ ions. The value of $\theta_{P}$~=~ -15.19~K is also considerably higher indicating the presence of  strong hybridization. An upturn in the low temperature heat capacity  gives an indication that the system may order magnetically just below the lowest temperature of our heat capacity measurements (0.4~K). The structure contains two sites for Yb ions and the present investigation suggests that Yb may be trivalent in one site while it may be significantly lower (close to divalent) in the other.
\vskip 1truecm 
\noindent 
Ms number ~~~~~~~~~~~~PACS number:~72.10.Fk, 72.15.Qm, 75.20.Hr, 75.30.Mb\\
\end{abstract}
\maketitle
\newpage
\section{Introduction}
\label{sec:INTRO}
\noindent
As a part of our continuing studies of the magnetic, electronic and transport properties of ternary 
rare-earth (R) intermetallic compounds of the type R$_2$T$_3$X$_5$ where T is a transition metal and X is 
an s-p element, we have recently focused our attention to the search for new Yb based compounds of the 235 type with interesting physical properties. In fact there are very few Yb based compounds of the type R$_2$T$_3$X$_5$ 
known due to the difficulty in preparation and characterization of these compounds. We had recently \cite{r1} synthesized high quality single phase samples of the compound Yb$_2$Fe$_3$Si$_5$ and established it to be a  Kondo-lattice compound  with heavy fermion behavior (among the highest $\gamma$ values~=~.5~J/Yb $mol~K^2$, for Yb compounds) and antiferromagnetic ordering at low temperatures. Pressure studies are in progress to look for superconductivity in this compound given that both the Tm and Lu compounds of the same series show superconductivity at low temperatures \cite {r2,r3,r4,r5}. Hence it is of interest to search for new Yb compounds with a similar structure because one can expect to see similar interesting physical properties. There have been some investigations on the compound Ce$_2$Ir$_3$Ge$_5$ \cite {r6,r7,r8} which  forms in U$_2$Co$_3$Si$_5$ structure, a structure closely related to the Yb$_2$Fe$_3$Si$_5$ structure. It has been shown to be a Kondo-lattice system with antiferromagnetic ordering and moderate heavy fermion behavior. The electron-hole analogy between Ce and Yb makes the compound Yb$_2$Ir$_3$Ge$_5$ an interesting candidate to investigate.

In this paper we report a comprehensive study of the structure, electrical resistivity, magnetic susceptibility 
and heat capacity of the new compound Yb$_2$Ir$_3$Ge$_5$. The susceptibility measurements indicate a mixed valent nature for the Yb ions while the low temperature resistivity follows a typical Kondo lattice behavior
and the heat capacity data shows an upturn below 5~K giving a hint that the compound may order magnetically
below the lowest temperature of measurement of 0.4~K. Such unusual properties have been reported earlier \cite {r9,r10} only  for certain Ce compounds of the form Ce$_2$T$_3$X$_9$ where T~=~ Rh,Ir with X~=~ Al,Ga. We find that  Yb$_2$Ir$_3$Ge$_5$ compound also shows such unusual properties where the magnetic response shows a typical mixed valent behavior while the low temperature resistivity and heat capacity data indicate heavy fermion or Kondo-lattice and magnetic ordering type of 
response. Yb$_2$Ir$_3$Ge$_5$ forms in the PMMN structure which allows the possibility of 2 sites for Yb ion. In this compound it is possible for Yb to have different valency in different sites. 
The paper is organised as follows. We first describe the details of sample preparation and characterisation and the measurement techniques in the section on experimental details. We then describe one by one the results and analysis of the magnetic susceptibility data, the resistivity data and the heat capacity data in subsequent sections. We then discuss 
the combined picture provided by all the measurements in the section on discussion. In the final section on conclusion we summarise all the results for the sample.
 
\section{EXPERIMENTAL DETAILS}
\label{sec:EXPT}
\noindent
A polycrystalline sample of Yb$_2$Ir$_3$Ge$_5$ was prepared by arc melting. First the binary Ir$_3$Ge$_5$
was prepared by arc melting the constituent elements taken in stoichiometric proportion on a water-cooled hearth in high purity argon atmosphere. The resultant button was flipped over and remelted 6 times to ensure homogenous  mixing. The weight loss in this stage was negligible. The binary was crushed into a fine powder and small pieces of Yb (with 5\% extra Yb) were mixed with it. The whole material was pelletized by a press into a closely packed tablet. This tablet was then arc melted and remelted only four times to make a compromise between Yb loss and homogenous mixing. We found that by pressing the Yb along with already prepared binary powder we could reduce the Yb evaporation considerably. No attempts to anneal the sample were made. A small piece of the as cast sample was used for powder X-ray diffraction. The crystal structure of the given sample was refined using {\bf FULL PROF} \cite{r11} and it is shown in Fig.~1. The slight difference in the intensity profile is probably associated with preferred orientation effects, which have not been taken into account in the present analysis. The refinement yielded the following values for the lattice constants: a~=~18.9474~\AA, b~=~15.4697~\AA and c~=~4.5960~\AA

The temperature dependence of the dc magnetic susceptibility ($\chi$) was measured using
the commercial Squid magnetometer (MPMS5, Quantum Design, USA) in a field of 1~kOe in the 
temperature range from 1.8 to 300~K. The resistivity was measured using a 
four-probe dc technique with contacts  made using silver paint on a bar shaped sample of 
1~mm thick, 10~mm length and 2 mm width cut from the arc melted ingot. The temperature was measured using a calibrated Si diode (Lake~Shore~Inc., USA) sensor. The heat-capacity in zero field between 0.4 to 35~K was 
measured using an automated adiabatic heat pulse method. A calibrated germanium resistance
thermometer (Lake~Shore~Inc, USA) was used as the temperature sensor in this range.
\section{RESULTS}
\subsection{Magnetic susceptibility studies}
\label{sec:SUS1}
\noindent
The temperature dependence of the inverse dc magnetic susceptibility
(1/$\chi_{dc}$) of polycrystalline Yb$_2$Ir$_3$Ge$_5$ in a
field of 1 kOe from 1.8 to 300~K is shown in the Fig.~2. with the inset showing the low temperature
$\chi$(T) data. The high temperature susceptibility (100~K$<$T$<$300~K) was fitted to a
modified Curie-Weiss expression which is given by,
$$ \chi~=~\chi_0~+~{C \over (T~-~\theta_p)}~, \eqno(1) $$
Here, $\chi_0$ is the temperature independent susceptibility, including the core-electron diamagnetism and the Pauli paramagnetism and the Van-Vleck terms. C is the Curie constant and $\theta_p$ is the Curie-Weiss temperature. The effective moment can be calculated from the value of the Curie constant from the relation,
$$ C~=~{\mu_{eff}^2~x \over 8}~, \eqno(2) $$
where $x$ is the concentration of Yb ions (x~=~2 for Yb$_2$Ir$_3$Ge$_5$).
The value which we get for  $\chi$$_0$ is (1.65$\times$10$^{-3}$ emu/mol) which is quite small. The value of the effective magnetic moment ($\mu_{eff}$) was found to be 3.66 $\mu_B$ and the value for the Curie-Weiss temperature $\theta_P$ was found to be $-$15.19~K. The estimated effective moment of only 3.66$\mu_B$ is much smaller than the value 4.54 $\mu_B$ for magnetic Yb$^{3+}$ ions.
This indicates that Yb may possibly be of mixed valent nature in this compound. However, usually for mixed valent compounds, one finds large and positive values of $\theta_P$ whereas we find a small and negative value ~=~-15.19~K for $\theta_P$ which is typical of compounds which undergo antiferromagnetic ordering at low temperatures. It is possible that the susceptibility would follow a Curie-Weiss law with an effective moment closer to the theoretical Yb$^{3+}$ value only at much higher temperatures above 300~K and crystalline electric fields are strong enough even at temperatures of the order of 100-300~K. 
We have also analyzed the data after subtracting the susceptibility of the Pauli paramagnetic Lu$_2$Ir$_3$Ge$_5$ sample.
The value of either $\mu_{EFF}$ or $\theta_P$ did not change significantly. It is also noteworthy that we do not observe any broad hump in the susceptibility typical of mixed valent systems.
There is a strong deviation of the data from the Curie-Weiss law below 100~K. This  could be due to the combined result of the Kondo effect and crystalline electric fields. Such a conjecture is substantiated
from the resistivity and heat-capacity studies described below. We do not observe any indication of magnetic ordering down to 1.8~K from the magnetization measurements.
\subsection{Resistivity studies}
\label{sec:RES1}
\noindent
The resistivity ($\rho$) vs temperature data of Yb$_2$Ir$_3$Ge$_5$ from 0.4~K to 300~K
is shown in Fig.~3. The room temperature resistivity value is about 200$\mu$$\Omega$-cm which 
is quite common among ternary rare earth silicides and germanides. The compound shows
metallic conductivity with some negative curvature of $\rho$(T) below about 150 K, which may 
result from crystal field interactions and/or s-d interband scattering of conduction
electrons. At about 10 K the resistivity goes through a broad shallow minimum and then slightly increases towards lower temperatures (see the inset of Fig. 4). However, below 2~K, the resistivity 
starts to decrease again as in the case of typical Kondo lattice compounds on the onset of coherent scattering. The latter hypothesis seems supported by the logarithmic character of the resistivity rise below 8~K (inset of Fig. 4). As is apparent from Fig. 4, the  temperature variation of the  resistivity of Yb$_2$Ir$_3$Ge$_5$ in the temperature range 2 K to 300 K can be well described by the formula,
$$ \rho(T)~=~\rho_0~+~\rho_0^i~+~4~R~\theta_D~({T\over{\theta_D}})^5\int^{\theta_D/T}_0
{x^5~dx\over{(1-exp(-x))(exp(x)-1)}}-K~T^3~-C_K~lnT ~ \eqno(3) $$ 
which comprises the residual scattering of conduction electrons on defects, impurities and spin-disorder
($\rho_0$+$\rho_0^i$), the Bloch-Gr\u{u}neisen scattering on phonons (third term), the
Mott s-d scattering (fourth term) and the Kondo scattering (fifth term). The least squares fitting procedure yields the following values of the parameters: $\rho_0$+$\rho_0^i$~=~126.9 $\mu$$\Omega$~cm, 
R~=~0.29 $\mu$$\Omega$~cm/K, $\theta_D$~=~110.5 K, K~=~4.8$\times$10$^{-7}$$\mu$$\Omega$~cm/K$^3$ and
c$_K$~=~0.32 $\mu$$\Omega$~cm. The non-negligible value of C$_K$ suggests that a weak Kondo effect
is operative at low temperature. The data could be fitted only up to 2~K below which the 
coherence sets in.
\subsection{Heat-capacity studies on Yb$_2$Ir$_3$Ge$_5$}
\label{sec:HCND}
\noindent
The heat capacity C$_p$ data of Yb$_2$Ir$_3$Ge$_5$ in the temperature range 0.4~K to 35~K 
is shown in the top panel A of Fig.~4. In this figure we have plotted the total heat capacity $C_P$, 
the heat capacity of iso-structural Lu$_2$Ir$_3$Ge$_5$ to estimate the lattice contribution $C_{lattice}$ 
to the total heat capacity and the magnetic heat capacity $C_{mag}$ calculated by subtracting the 
heat capacity of Lu$_2$Ir$_3$Ge$_5$ from the total heat capacity. 
We have taken care of the mass difference between Yb and Lu while estimating the lattice contribution for the Yb sample. The low temperature heat capacity shows an upturn below 6~K and starts increasing as we go down in 
temperature. Panel B shows the data plotted as $C_P$/~T vs $T^2$. This plot emphasizes the sharp 
upturn in the data at low temperatures. The magnitude and sharpness of the upturn as seen in the 
inset panel I of panel B strongly indicates a possible magnetic ordering on going further down in temperature. 
The inset II of the panel B shows the temperature dependence of the  estimated magnetic entropy S.
The entropy reaches a value of about .27~Rln(8) at 35~K which is much lower than the value 
$R$ln(2$\bf J$+1) with $\bf J$~=~7/2 for Yb. Thus the total entropy has not been released 
even upto a temperature of 35~K. This indicates a significant contribution from crystal field effects. 
The magnetic contribution to the heat capacity $C_{mag}$ plotted in the panel A shows a broad hump peaked
around 27~K indicating the presence of a Schottky anomaly.
The Schottky type anomaly at higher temperatures and the upturn in heat capacity at low temperatures 
makes it impossible to fit the data in any temperature range to the equation 
$\gamma$T + $\beta$$T^3$ to get a measure of the Sommerfeld constant $\gamma$.
\section{DISCUSSION}
\label{DIS}
Let us now try to explain why the combined data of all the measurements described above is unusual before we give arguments, which attempt to understand such a behavior.
The nature of the ground state (GS) in such systems where an array of 4f moments is embedded in a metallic environment results from a competition between two energy scales. The onsite Kondo effect (favoring a non-magnetic GS) with a characteristic energy given by the Kondo temperature T$_K$ and the intersite RKKY (characterized by T$_{RKKY}$)interaction favoring a magnetic GS. The GS physics depends on the final balance between these two mechanisms which is driven by the strength of the hybridization between the localized f electrons and the itinerant conduction electrons. In general T$_K$ values are smaller for heavy fermion compounds which display Kondo lattice behavior with a large $\gamma$ and a fall in $\rho$ due to coherent
scattering. The effective moment of the 4f ions (estimated from the high temperature $\chi$ data) will be nearly equal to its theoretical value. However, one can see from our data that the Curie-Weiss fit to the high temperature inverse susceptibility gives an effective moment considerably reduced from the Yb$^{3+}$ value suggesting that Yb may be in a mixed valent state in this compound whereas the low temperature transport and heat capacity measurements show a Kondo lattice type of behavior with strong indications of a possible magnetic ordering just below 0.4~K. Although the calculated effective paramagnetic moment is quite low, the overall susceptibility behavior looks more like that of a system which is largely trivalent rather than one which is mixed-valent [e.g., compare the magnetic susceptibility of CeNiGe$_2$ or CeNiSn$_2$ (fig.6 of \cite{r12}) in which Ce is in the 3$^+$ state with that of CeNiSi$_2$ (fig.4 of \cite{r12}) in which Ce is mixed-valent].
One can try to explain the behavior of Yb$_2$Ir$_3$Ge$_5$ if we realize that it forms in the PMMN structure which allows Yb to have two inequivalent sites. Hence it is possible for Yb ions to have a different valency at the two different sites. We believe that this is infact the case and Yb on one of the sites has a valency close to the 2$+$ while at the other site it has a valency of 3$+$. In such a case the non-magnetic Yb$^{2+}$ ions would exhibit Pauli paramagnetism while the magnetic Yb$^{3+}$ ions would follow a Curie-Weiss kind of behavior. Thus one would see a reduced effective moment depending on the relative occupancies of the two different sites. The Yb$^{3+}$ ions would also be the ones participating in the Kondo effect and may undergo magnetic ordering at low temperature. One needs to have single crystal x-ray data to calculate the exact positions and occupation factors of Yb and other elements in the unit cell to get a clearer understanding of the underlying physics.
\section{CONCLUSION}
\label{sec:CON}
To conclude, we have synthesized a new Yb based compound namely Yb$_2$Ir$_3$Ge$_5$.
which forms in the orthorhombic structure with space group PMMN unlike its Ce analogue which 
forms in the tetragonal U$_2$Co$_3$Si$_5$ type structure. Iso-structural Lu$_2$Ir$_3$Ge$_5$ was also 
synthesized to estimate the lattice contribution to the heat capacity. It is also found to take the same structure as the Yb compound.
A Curie-Weiss fit to the high temperature (100-300~K) inverse magnetic susceptibility data gives 
an effective moment of 3.66$\mu_B$ which is far less than 
the value 4.54$\mu_B$ for magnetic Yb$^{3+}$ ions indicating that Yb may be in a mixed valent state in this compound. However, the structure PMMN can have two sites for Yb ions and the valency of Yb may be different
from 3 in one of the sites thus giving a net reduced effective paramagnetic moment. Such behavior has been observed in some Cerium based compounds with multiple
Ce sites \cite{r13}. This is the first report to our knowledge on an Yb based compound showing such behavior. Our resistivity measurements show that Yb$_2$Ir$_3$Ge$_5$ behaves like a Kondo-lattice below 2~K.
The heat capacity data shows signs of a strong CEF influence with a Schottky type anomaly around 27~K.
Finally we could not detect any magnetic ordering in the compound down to 0.4~K though the low 
temperature heat capacity indicates that the compound may possibly undergo magnetic ordering below 0.4~K.

\newpage
\begin{figure}
\caption{Observed and fitted X-ray power diffraction pattern of
Yb$_2$Ir$_3$Ge$_5$ using FULL PROF program. The vertical bars $-----$
mark the Bragg peaks and the difference between observed and calculated
intensity is also plotted below.
\label{Xray}}
\end{figure}
\begin{figure}
\caption{Temperature dependence of the inverse susceptibility (1/$\chi_{dc}$) of
Yb$_2$Ir$_3$Ge$_5$ in a  field of 1~kOe from 1.8 to 300~K.  The inset shows the low temperature
$\chi$ behavior. The solid line is a fit to the Curie-Weiss relation (see text).
\label{fsus1}}
\end{figure}
\begin{figure}
\caption{Temperature dependence  of the resistivity ($\rho$) of
of Yb$_2$Ir$_3$Ge$_5$ from 1.8 to 300~K. The inset shows
the $\rho$ data from 1.8 to 18~K on an expanded scale. 
The solid line shown in the inset as well as in the main figure 
is a fit to the model, which includes contribution from the Kondo effect.
(see text for details).
\label{fres1}}
\end{figure}
\begin{figure}
\caption{Plot of the heat-capacity (C$_p$) vs temperature (T) of
Yb$_2$Ir$_3$Ge$_5$  from 0.4 to 35~K. The top panel A shows the total
heat capacity C$_p$, the lattice heat capacity C$_{lattice}$, the magnetic
contribution C$_{mag}$ to the heat capacity. The lower panel B shows the C$_P$/T vs T$^2$ data. 
The inset I of panel B describes the sharp rise at very low temperatures while the inset II 
of the same panel depicts the temperature dependence of magnetic entropy S.
\label{fcp1}}
\end{figure}
\end{document}